\documentclass[9pt,twocolumn,twoside]{osajnl}
\usepackage{soul}
\journal{ol} % Choose journal (ao, aop, josaa, josab, ol, optica, pr)

\setboolean{shortarticle}{true}
\title{Fully reconfigurable coherent optical vector-matrix multiplication}

\author[1 $\dagger$]{James Spall}
\author[1,2 $\dagger$]{Xianxin Guo}
\author[1]{Thomas D.\ Barrett}
\author[1,3 *]{A. I. Lvovsky}

\affil[1]{University of Oxford, Clarendon Laboratory, Parks Road, Oxford  OX1 3PU, UK}
\affil[2]{Institute of Fundamental and Frontier Sciences, University of Electronic Science and Technology of China, Chengdu, Sichuan 610054, China}
\affil[3]{Russian Quantum Center, Skolkovo, 143025, Moscow, Russia}

\affil[*]{Corresponding author: alex.lvovsky@physics.ox.ac.uk}

\doi{\url{}}

\usepackage{amsmath}
\usepackage{bm}
\usepackage[normalem]{ulem}
\usepackage{siunitx}

\begin{abstract}
Optics is a promising platform in which to help realise the next generation of fast, parallel and energy-efficient computation. We demonstrate a reconfigurable free-space optical multiplier that is capable of over \num{3000} computations in parallel, using spatial light modulators with a pixel resolution of only $340\times340$. This enables vector-matrix multiplication and parallel vector-vector multiplication with vector size of up to 56. Our design is the first to simultaneously support optical implementation of reconfigurable, large-size and real-valued linear algebraic operations. Such an optical multiplier can serve as a building block of special-purpose optical processors such as optical neural networks and optical Ising machines.
\end{abstract}

\setboolean{displaycopyright}{False}

\begin{document}

\maketitle

%%%%%%%%%%%%%%%%%%%%%%%%%%  body  %%%%%%%%%%%%%%%%%%%%%%%%%%

\section{Introduction}

    Matrix multiplication is a core element in a wide range of computational problems, ranging from image processing to machine learning. For example, modern neural networks rely on learning models with tens of millions of parameters, for which the predominant computational cost is a correspondingly huge number of matrix multiplications. This remains a significant computational bottleneck, even with the development of bespoke digital hardware such as graphics and tensor processors. 
    
    Matrix multiplication consists of evaluating pairwise products of large arrays of numbers, followed by addition of all the products within a particular row of column. If the matrix elements are encoded in optical signals,  these operations can be processed in parallel thanks to the  coherence and superposition properties of light. Hence optics offers a promising analogue platform to realise the next generation of processors capable of fast, parallel and power-efficient linear algebraic operations \cite{Caulfield2010}.
    
    Multiplication of matrices can be viewed as a set of vector-matrix multiplications (VMMs) implemented in concert. Various optical VMM (OVMM) systems have been proposed and demonstrated in both free space \cite{Goodman1978, Tamura1979, Farhat1985, Mosca1989, Wu1994} and integrated photonic circuits \cite{Gruber2000, Harris2017, Shen2017}. Most free-space implementations are based on the Stanford multiplier design \cite{Goodman1978}, where input vector values are encoded in the intensity of a horizontal array of incoherent light sources. Cylindrical and spherical lenses are used to vertically spread each vector element onto one column of an amplitude mask, such as a photographic transparency, which encodes the matrix values. A second set of lenses is then used to converge each row onto a vertical photo-sensitive array, completing the multiplication.
    	
    However, by encoding values in the intensity of incoherent light, the Stanford multiplier can only multiply positive real values. To handle negative or complex values, the multiplication must be decomposed and requires multiple optical setups, which increases the computational cost. In 1979, Tamura et al.~developed a coherent OVMM system to process real-valued operations \cite{Tamura1979}. However, technology of the time prevented reconfigurability and restricted the practical demonstration to binary matrices on photographic transparencies. 
	
	In recent years, reconfigurable on-chip OVMM has been achieved by utilising integrated optical interference units, consisting of Mach-Zehnder interferometers and phase shifters \cite{Shen2017}. A real-valued vector is encoded in the electric field of coherent light, and a real-valued matrix is realised through a large network of interference units. Such integrated OVMM systems is able to not only process real values, but also be dynamically reconfigured via thermo-optical or electro-optical phase shifters. Yet, given current technology, matrix sizes of only $4 \times 4$ have been demonstrated in practice \cite{Shen2017}. With this architecture, realising arbitrary matrices of larger size is quadratically expensive in the number of required phase shifters. 
	
	Here we demonstrate a reconfigurable coherent optical multiplier with matrix size up to $56\times56$, an order of magnitude larger than achieved previously \cite{Tamura1979}. Our free-space system is capable of performing VMM, as well as parallel vector-vector multiplication (p-VVM), and our results are obtained using spatial light modulators (SLMs) with only $340\times340$ pixels. With the achievable resolution ultimately determining the maximum matrix size, current off-the-shelf models could realise over $10^5$ parallel computations and matrix sizes of several hundred.
	
\section{Concept of coherent OVMM}

	\begin{figure}[ht]
		\centering
		\includegraphics[width=\linewidth]{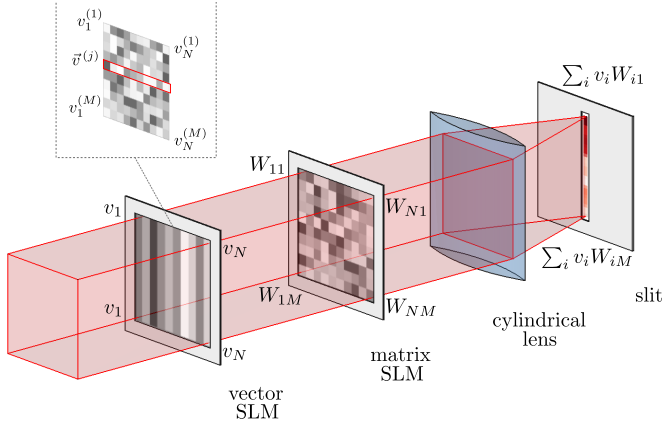}
		\caption{Conceptual diagram of coherent OVMM using two ideal spatially variable absorbers, cylindrical lens and slit. Encoding multiple input vectors in the vector SLM, as shown in the inset diagram, enables  parallel vector-vector multiplication.}
		\label{fig: ovmm schematic}
	\end{figure}

    We begin by outlining, in principle, how to perform the real-valued calculation $\vec{u} = \bm{W} \vec{v}$ with optics, where $\bm{W}$ is an $M\times N$ matrix, and $\vec{u}, \vec{v}$ are vectors of sizes $M$ and $N$, respectively. 
    Figure \ref{fig: ovmm schematic} shows a conceptual diagram. Our aim is to encode $\vec{v}$ and $\bm{W}$ in the electric field of a coherent light source to perform element-wise multiplication between the vector and each row of the matrix, then summation along each row. 
    
   For this conceptual explanation, we treat the SLMs as ideal devices capable of arbitrarily modulating the amplitude and phase of the electric field at any transverse position. These modulators encode the vector values, expanded along each column such that $v_{ij} {=} v_i\ \forall\ j\in\{1,\ldots,M\}$, and matrix elements, $W_{ij}$, as the field transmission. In general, these vector and matrix elements can be complex-valued, but here we restrict them to real-valued. A coherent light source produces a uniform collimated electric field $E_0$, which propagates through both SLMs, resulting in a field with the spatial profile
    \begin{align}
    \label{eq:ImagePlane}
    E_{ij} = E_0 v_i W_{ij},
    \end{align}
    equivalent to element-wise multiplication between the vector and each row of the matrix. The field is then focused in the horizontal dimension by a cylindrical lens onto a narrow vertical slit. The lens performs a Fourier transform in that dimension, and the slit selects the zero spatial frequency component --- that is, the sum of $E_{ij}$ over each row.  Therefore the field immediately after the slit,
    \begin{align}
    \label{eq:Eout}
    E_{\text{out},j} & =E_0 \sum_{i=1}^N v_i W_{ij},
    \end{align}
    encodes the desired VMM result, up to some global scaling factor.
    
    This setting can be generalized beyond VMM. We can encode $M$ different vectors  $ \left\{ \vec{v}^{\,(1)} \dots \vec{v}^{\,(M)} \right\}$ in different rows of the the first SLM, and interpret the matrix encoded on the second SLM as another set of $M$ vectors, $\bm{W} = \left[ \vec{w}^{\,(1)} \dots \vec{w}^{\,(M)} \right]^T$. In this case, the $j^{\text{th}}$ element of the output field \eqref{eq:Eout} represents the inner product $ \vec{v}^{\,(j)} \cdot \vec{w}^{\,(j)} $, so our setup multiplies $M$ pairs of vectors of size $N$ in parallel. We note that VMM is simply a particular case of p-VVM when all vectors $\vec{v}^{\,(j)}$ are the same.

\section{Methods}

	\begin{figure}[t]
		\centering
		\includegraphics[width=\linewidth]{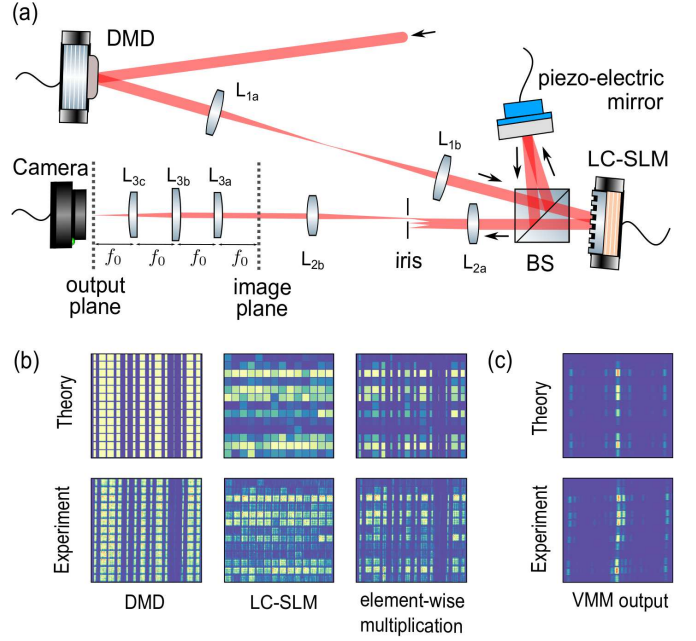}
		\caption{\textbf{(a)} Experimental schematic of the reconfigurable coherent OVMM. The vertical $4f$ imaging cylindrical lenses $L_{3a}$ and $L_{3c}$ have focal lengths of $f_0=200$ mm, and the horizontal Fourier transform lens $L_{3b}$ has focal length of $f=2f_0=400$ mm. \textbf{(b)} Theoretical and experimental images of the fields individually created by the DMD and LC-SLM, as well as their element-wise product for one OVMM example, recorded with the camera in the image plane. \textbf{(c)} Theoretical and experimental OVMM result acquired in the output plane.}
		\label{fig: experimental schematic}
	\end{figure}

    The experimental schematic is shown in Fig.~\ref{fig: experimental schematic}(a). We use a digital micromirror device (DMD) and liquid-crystal spatial light modulator (LC-SLM) to encode the vector and matrix respectively. The reason for the choice of a DMD for vector encoding is a faster update rate, which is desired for many applications such as optical neural networks and Ising machines \cite{Inagaki2016, McMahon2016, conti2019,guo2019,Shen2017,Wang2018,Mercante2018}. The trade-off is that the DMD can only output positive vectors. A further shortcoming is that each of its pixels can take on a binary value, limiting the relative precision with which a vector element can be encoded, to the number of DMD pixels representing it. For applications requiring vectors with higher-precision signed values, the DMD can be replaced by a second LC-SLM without conceptual changes to the setup.  
    
    The DMD and LC-SLM models are TI Discovery 1100 and Meadowlark Optics Model P512, and each vector or matrix element is encoded by a block of pixels. The size of the block is chosen  dependent on the matrix size, but was at least $3\times3$ on the DMD and $5\times5$ on the LC-SLM. The matrices we encode are square of size $N\times N$, but the last row of both matrices is used for reference as described below, so we have $M=N-1$.

    For the DMD, each pixel is a binary `on/off' state, and the encoded value is proportional to the number of pixels switched `on' within each block.  We illuminate the DMD with a large-waist laser beam, and the DMD plane is imaged to the LC-SLM by a $4f$ imaging system comprising a pair of spherical lenses $L_{1a}$ and $L_{1b}$. The LC-SLM is a phase-only modulator, so the control of both the amplitude and phase of the reflected field is achieved by  modulating the phase shifts in a grating-like pattern. The offset and amplitude of this grating determine the phase and amplitude of the field in the first diffraction order \cite{Arrizon2007}. A spatial filter, comprising two spherical lenses $L_{2a}$ and $L_{2b}$ and an iris, is then applied to eliminate other diffraction orders. The LC-SLM is calibrated to correct for local variation in pixel phase response and a global curvature of both the LC-SLM and DMD \cite{pushkina2020}.
   
    The field after the spatial filter (marked in Fig.~\ref{fig: experimental schematic}(a) as ``image plane") then represents the element-wise product \eqref{eq:ImagePlane} of $\vec v$ and $\bm{W}$. We use a cylindrical lens $L_{3b}$ to perform a Fourier transform in the horizontal $x$ direction, and two additional cylindrical lenses $L_{3a}$ and $L_{3c}$ to perform $4f$ imaging in the vertical $y$ direction. A CMOS camera (UI223-SE-M, $1024\times768$ pixels) measures the entire field at the output plane after these cylindrical lenses. The central strip of the image with a width of 32 camera pixels contains the VMM output. We  implement the slit shown in Fig.~\ref{fig: ovmm schematic} digitally by acquiring and processing the data from this area.
    
    To measure the (signed) real-valued output, we construct an interferometer with a beam splitter and a continuously scanning piezoelectric-driven mirror near the LC-SLM. The amplitude and phase of all rows are obtained from fitting a sinusoidal curve to the interference fringes recorded while scanning the piezoelectric-driven mirror. Throughout the experiment, we fix one row ($j=N$) of the DMD and LC-SLM patterns with uniform, constant values, to serve as the reference row. If the output vector element is in (out of) phase with the reference row, we determine it as a positive (negative) value. We note that our measurement approach is just one of many possible methods. For example, single-shot interferometric measurement, without the need for continuous scanning and numerical fitting, can be implemented using a phase-locked reference beam. The camera output is calibrated by setting  all the DMD and LC-SLM blocks to the values corresponding to $v_i=W_{ij}=1$.

    Fig.~\ref{fig: experimental schematic}(b,c) illustrates one example of OVMM. In the first two columns of Fig.~\ref{fig: experimental schematic}(b) we display the patterns individually produced by the DMD and LC-SLM, acquired by setting the respective other modulator to uniform maximal reflection and placing the camera temporarily into the image plane. The last column of Fig.~\ref{fig: experimental schematic}(b) shows the element-wise  product of the two patterns. In Fig.~\ref{fig: experimental schematic}(c) we demonstrate the action of the cylindrical lenses, showing theoretical and experimental images of the field at the output plane, which corresponds to the complete OVMM result. We see good agreement between experiment and theory, however the experimental output image shows some aberration, attributed to imperfectly correcting the wavefront curvature with the LC-SLM, and imperfect cylindrical lenses. The aberration is unchanged throughout the experiment, and we take the values along the curved central line as our OVMM result.

\section{Results}
	
	\begin{figure}[tbp]
		\centering
		\includegraphics[width=\linewidth]{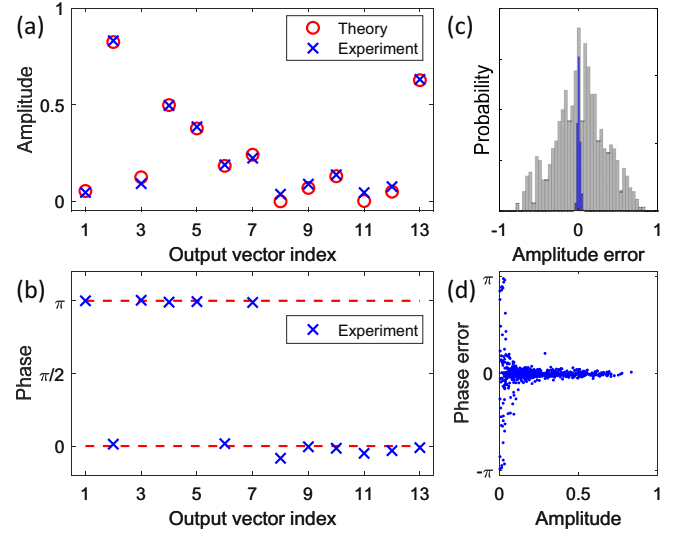}
		\caption{Multiplier output. \textbf{(a)} Amplitudes and \textbf{(b)} phases of one p-VVM example, with size $N=14$. \textbf{(c)} Error between the theoretically expected and measured amplitudes (blue histogram) as well as the distribution of theoretically expected products (grey histogram) for 50 random examples, that is, 650 data points. \textbf{(d)} Phase error distribution as a function of the output amplitude.}
		\label{fig: VMM example}
	\end{figure}

	We characterise our system using p-VVM, as this is a more general task which includes VMM as a particular case. It is also more challenging experimentally, as the system is more sensitive to misalignment between the DMD and LC-SLM when the DMD is displaying many different vectors. For a fixed vector size $N$, we perform a batch of 50 p-VVMs, that is $50M=50 (N-1)$ vector-vector multiplications in total (the subtraction of 1 accounts for the reference row). All DMD vector values $v_i^{(j)}$ are randomly chosen in the range $[0,1]$, all LC-SLM values $w_i^{(j)}$ in the range $[-1,1]$, and every vector is unique. The amplitude of the output vector is normalised by the maximum possible output of the reference row, so that the output amplitude falls in the range $[0,1]$, and the output phase is measured in the range $[-\pi,\pi]$.

	Fig.~\ref{fig: VMM example}(a,b) shows the measured amplitude and phase of one p-VVM output with $N=14$. For each of the $(N-1)\times 50 =650$ multiplications, we take the difference between the experimental results and theory values, to find the error distributions in amplitude and phase shown in Fig.~\ref{fig: VMM example}(c,d). The standard deviation between the measured and theoretically expected amplitudes [blue histogram in Fig.~\ref{fig: VMM example}(c)] is $\sigma_A=0.0174$. For comparison, we also plot the distribution of theoretically expected p-VVM outputs for the chosen set of 50 vectors as the grey histogram. The standard deviation of this distribution is  $0.3166$, a factor of $18.2$ higher than $\sigma_A$. 
    
    Fig.~\ref{fig: VMM example}(d) shows the phase errors as a function of the output amplitude. Although the distribution concentrates around zero, larger phase errors occur at lower output amplitudes due to poorer fitting of the interference patterns. Considering only points with non-negligible amplitude of $A > 2\sigma_A$ (i.e.\ those for which a phase error will practically affect OVMM performance), the phase  standard deviation is $\sigma_{\phi} = 0.0646 \pi$.

\begin{figure}[tbp]
		\centering
		\includegraphics[width=\linewidth]{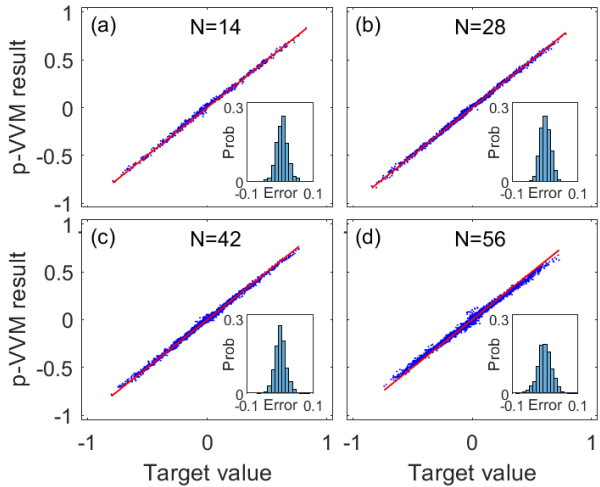}
		\caption{Comparison between experimental p-VVM result and target values with vector sizes of 14, 28, 42 and 56 in \textbf{(a)}, \textbf{(b)}, \textbf{(c)} and \textbf{(d)}, respectively. Experimental data (blue dots) are collected from 50 parallel p-VVM examples at each size, and the expected target values are plotted as theory lines (red lines). Insets are the corresponding real-valued error distributions.}
		\label{fig: VMM with different sizes}
	\end{figure}

	Having measured the amplitude and phase of each multiplication result, we convert them to real numbers in the range $[-1,1]$, by rounding the phase to 0 or $\pi$. We then analyse the precision of the multiplier over the real values for dimensions 14, 28, 42 and 56, the results of which are shown in Fig. \ref{fig: VMM with different sizes}. The error distributions (insets in Fig.~\ref{fig: VMM with different sizes}) have standard deviations $\sigma_N$ of $0.0170$, $0.0155$, $0.0169$ and $0.0237$ for $N=14$, 28, 42 and $56$, respectively. The corresponding signal-to-noise ratios, defined as the ratio of the p-VVM output distribution width and $\sigma_N$, are calculated to be $18.6$, $19.5$, $16.5$ and $10.2$.

	The most significant factor determining the multiplier performance is the accuracy of the LC-SLM calibration, which is independent of the multiplication size. However, at size $N=56$, each LC-SLM vector element is only modulated by just one grating period. Therefore, the first diffraction order begins to distort, and we have larger phase error leading to imperfect constructive or destructive interference between vector elements. In this case, the maximum experimental amplitude tends to be smaller than the theory value, as can be seen from Fig.~\ref{fig: VMM with different sizes}(d). Interestingly, we also see the amplitude errors to be higher for lower amplitudes. This is because, for lower absolute values of $W_{ij}$, the reflection of the SLM into the first diffraction order is lower, meaning that more light is emitted into other diffraction orders, resulting in higher background noise.

\section{Discussion}
	
There is a clear path to improving the multiplier performance to the level comparable to, or even above that of CPU/GPUs with our approach. The system throughput and computation speed can be enhanced by increasing modulator resolution, bandwidth, and performing single-shot interferometric measurement.
	
Our LC-SLM model has $512 \times 512$ pixels, but due to unresponsive pixels and large surface curvature at the edges, the working area is reduced to approximately $340 \times 340$ pixels. In our encoding method we use a grating period of four pixels, with a minimum block size of $5\times5$ pixels and a one-pixel gap between blocks. We can therefore achieve a maximum matrix dimension of 56. Current DMD and LC-SLM technology provides pixel areas up to approximately $2000 \times 1000$. This corresponds to a maximum matrix dimension over $300\times150$ for our setup, or $\sim 10^5$ operations in a single VMM.  With existing commercial DMD models with pattern refresh rates of up to \SI{20}{\kilo\Hz}, the multiplier can perform $10^9$ operations per second. In the future, this rate can be increased by another five orders of magnitude with GHz speed electro-optic modulator arrays \cite{Wang2018,Mercante2018},surpassing current electronic computing speed of $10^{12}$ operations per second. Note that free-space optical neural networks exhibiting this performance level, albeit not utilizing a VMM such as ours, have recently been reported \cite{Miscuglio2020,Zhou2020}.
	
It is also interesting to estimate the energy efficiency of free-space OVMM. Assuming the average output power of $10^{-8}$ Watt per mode, which can be easily measured with commercially available photodetectors, the power consumption of a multiplier with $10^2$ output modes is $10^{-6}$ Watt. Our system has an overall efficiency of $7\%$, which includes $60\%$ efficiency of DMD, $30\%$ of LC-SLM and $40\%$ beam utilization efficiency. Taking into account this efficiency, as well as a productivity of $10^9$ operations per second, we find the energy consumption of $10^{14}$ operations per Joule, which outperforms current digital devices with $\sim10^{9}$  operations per Joule. The fundamental limit of energy efficiency is set by the required signal-to-noise ratio. To suppress the shot noise below 1$\%$ of signal level, $10^4$ photons are needed per output mode, or $10^6$ photons per VMM, yielding an energy efficiency limit of $\sim 10^{17}$ operations per Joule. 
	
However, even without further increases in performance available with current technology, our device already demonstrates a significant step forward. The computational power of our system already opens the door to  practical applications. This includes using optical Ising machines to find the ground state of Hamiltonians with hundreds of arbitrarily coupled spins \cite{Inagaki2016, McMahon2016, conti2019}, or demonstrating optical neural networks with hundreds of neurons in each layer where inference tasks can be carried out with nano-second latency \cite{guo2019, Shen2017}.

\section*{Acknowledgments}

A.L.’s research is partially supported by Russian Science Foundation (19-71-10092). X.G. acknowledges funding from the University of Electronic Science and Technology of China.

\medskip

\noindent\textbf{Disclosures.} The authors declare no conflicts of interest.

\medskip

\noindent $^{\dagger}$ These authors contributed equally to this work.

%%%%%%%%%%%%%%%%%%%%%%% References %%%%%%%%%%%%%%%%%%%%%%%%%

\bibliography{coherentVMM}

\bibliographyfullrefs{coherentVMM}

\end{document}